# Accuracy Improvement for Fully Convolutional Networks via Selective Augmentation with Applications to Electrocardiogram Data

**Lucas Cassiel Jacaruso[1]**

[1]University of Southern California, Los Angeles, CA 90007 USA [1]Universität für Musik und darstellende Kunst Graz, Mondscheingasse 7, 8010, Graz, Steiermark, Austria

Corresponding Author: Lucas Cassiel Jacaruso (personal email (preferred): jacaruso33@gmail.com, institutional email: [jacaruso@usc.edu](mailto:jacaruso@usc.edu))

Phone: 551-579-0801

Mailing Address: 52401 Double View Drive Suite 3453 Idyllwild CA 92549

# Abstract

Deep Learning methods have shown suitability for time series classification in the health and medical domain, with promising results for electrocardiogram data classification. Successful identification of myocardial infarction holds lifesaving potential, and any meaningful improvement upon deep learning models in this area is of interest. Conventionally, data augmentation methods are applied universally to the training set when data are limited to ameliorate data resolution or sample size. In the method proposed in this study, data augmentation was not applied in the context of data scarcity. Instead, samples that yield low confidence predictions were selectively augmented to bolster the model's sensitivity to features or patterns less strongly associated with a given class. This approach was tested for improving the performance of a Fully Convolutional Network. The proposed approach achieved 90% accuracy for classifying myocardial infarction as opposed to 82% accuracy for the baseline, a marked improvement. Further, the accuracy of the proposed approach was optimal near a defined upper threshold for qualifying low confidence samples and decreased as this threshold was raised to include higher confidence samples. This suggests exclusively selecting lower confidence samples for data augmentation comes with distinct benefits for electrocardiogram data classification with Fully Convolutional Networks.

# Keywords

Fully Convolutional Networks, Deep Learning, Time Series Classification, Electrocardiogram Data, Data Mining

# Introduction

A time series is defined as data that are sequentially indexed in time order, and are ubiquitous in nature and society. Time series classification (TSC) is one of data mining's persistent challenges. Applications of TSC abound in fields including agriculture, medicine, and engine prognostics [1-3], a common goal being to detect instances of sub-optimal behavior or decreased health (biologically or mechanically) as just one real-world example. Dozens of new TSC algorithms were introduced in the last four years alone [4,5]. This trend has been intensified by the increasing availability of real-world datasets. In fact, the classification of any inherently ordered data (temporally or otherwise) can be treated as a TSC problem [4,6], making for a vast breadth of real-world applications. Of particular interest in this paper is the healthcare application. Emergency departments often fail to detect high-risk electrocardiogram (ECG) findings, resulting in preventable casualties as shown by Masoudi et.al. [7]. Early detection of myocardial infarctions (heart attack due to cardiac ischemia) is one area where TSC models can have a material impact on patient outcomes with the potential to save lives.

It can be argued that the most standard baseline approach for TSC is the Nearest Neighbor (NN) classifier used alongside the Dynamic Time Warping (DTW) similarity measure [8,9]. This approach involves using DTW to calculate the similarity between samples based on the degree of nonlinear "warping" necessary to achieve the best alignment between the series. DTW with NN classification has proven difficult to outperform even with more sophisticated classification methods. [9,10]. Research has however continued to explore unique ensembling methods of NN classifiers to improve accuracy [11, 12].

Deep Learning methods have also gained considerable traction for TSC applications. Conventionally applied to computer vision [13,14], applications of CNNs and FCNs have since extended to areas including healthcare [15] and natural language processing [16,17]. CNNs and FCNs have proven successful for time series tasks including TSC [18, 19]. Deep learning models benefit from a large amount of training data, so time series data augmentation methods have gained interest for cases when data are limited. [20]. Even still, the literature does not extensively address the performance impact of time series augmentation on FCNs specifically within the health and medical domain.

The focus of this paper is to improve the performance of the FCN via a method predicated on several assumptions. The first assumption is that a deep learning classification model will gravitate towards the unique patterns in the data more strongly associated with a given class [5]. If a sample has many patterns or features strongly associated with a given class, the model's prediction on that sample will naturally be higher confidence. By extension, it is assumed the model will yield lower confidence predictions on test samples containing features more weakly correlated with a given class. The latter will simply be referred to as low confidence samples. This paper explores the possibility of introducing resampled segments of low-confidence samples (from an intermediate test set) back into the training set. Once the resampled segments have been reintroduced into the training set, the model is retrained then tested on a new test set to evaluate final performance. The rationale behind this approach is loosely inspired by human intuition. Although there is no official consensus, many authorities in psychology and cognitive science believe intuition is not altogether different from analytical reasoning but simply grapples with the less notable features of one's experience rather than the more obvious ones [21-23]. Reintroducing segments of low confidence samples to the training set (which have a stronger likelihood of containing the less prominent features/patterns of a given class) is intended to slightly bolster sensitivity to such features/patterns in a way roughly analogous to intuition. This parallel, however, is drawn cautiously. In this light, the accuracy of a modified implementation of the FCN is assessed in relation to a conventional implementation of the FCN as a baseline.

Typically, data augmentation is applied universally to a dataset when data are insufficient [20]. The practice of augmenting time series segments obtained solely from low confidence samples is yet unconventional and is the main novelty. The notion of selectively targeting low confidence samples for augmentation is virtually unexplored in the current literature about FCNs. Various adjustable parameters are inherent to the newly proposed approach including the threshold which defines low confidence predictions and the segment size to be resampled.

The proposed method is used to build classification models for heartbeat electrical activity. The dataset has two classes of example heartbeats, one being normal heart activity and the other a myocardial infarction (MI). Therefore, the classification task is binary. The performance impact of altering the parameters within the new method is evaluated across several example models. The rest of this paper is organized as follows. In section 2, related work is revisited, and background is provided on the specifics of FCN and existing time series augmentation/transformation methods. Section 3 describes the data used, the architecture of the proposed approach, and the experimental setup. Section 4 details the results of the new method and compares them to the baseline on the same data. In section 5, an interpretation of the results is provided along with potential limitations of the study. Finally, section 6 summarizes the findings and discusses future research directions.

# Background/Related Work

## Fully Convolutional Networks

First proposed by Wang et al. [24], the FCN has shown considerable efficiency and suitability for time series classification problems. In a conventional CNN, a filter is applied to the input data to extract a map of activations, or a feature map [25]. The filter (sometimes referred to as a kernel) shifts over regions of the data to extract features. The spatial sizes of convolved features are then reduced by a local pooling layer. The dimensionality reduction provided by local pooling improves computational efficiency [26]. The following equation describes a convolution for a centered timestamp t:

$$C_t = f(\omega * X_{t-l/2:t+l/2} + b) \mid \forall t \in [1, T]$$

where $C$ is the result of a dot product (*) on a time series $X$, which has a length $T$. $\omega$ is a filter of length $l$, $b$ is the bias parameter and $f$ is the Rectified Linear Unit (ReLU) activation function.

Intuitively, the result of convolution on an input univariate time series is a multivariate time series where the number of dimensions matches the number of filters.

For all timestamps, identical convolutions are performed, i.e. with the same filter value and bias parameter. This allows for a quality called weight sharing [27] which contributes the CNN's ability to learn invariant filters in the time dimension.

The FCN is essentially a CNN that omits the aforementioned local pooling layers, so that time series length stays consistent during convolutions. Features are instead fed from the convolution blocks to a global pooling layer [28] with a softmax layer serving as the final output layer. The softmax function returns a normalized probability distribution over classes (which always sums to 1), and is defined as follows:

$$\sigma(\vec{Z})_i = \frac{e^{Z_i}}{\sum_{j=1}^{K} e^{Z_j}}$$

Where $\vec{Z}$ is the input vector, $e^{Z_i}$ the exponential function for the input vector, $K$ is the number of classes, and $e^{Z_j}$ is the exponential function for the output vector.

Global pooling takes the average of each feature map and directs the resulting vector to the softmax layer, which produces the final label. In other words, the time series is aggregated across the whole dimension rather than a sliding window (as would be the case with local pooling) [5]. Zhou et al. [29] demonstrate the advantages of global pooling, which include the reduction of overfitting. Each block consists of the convolutional layer, a batch normalization layer (to expedite convergence) [30], and a ReLU activation layer [31].

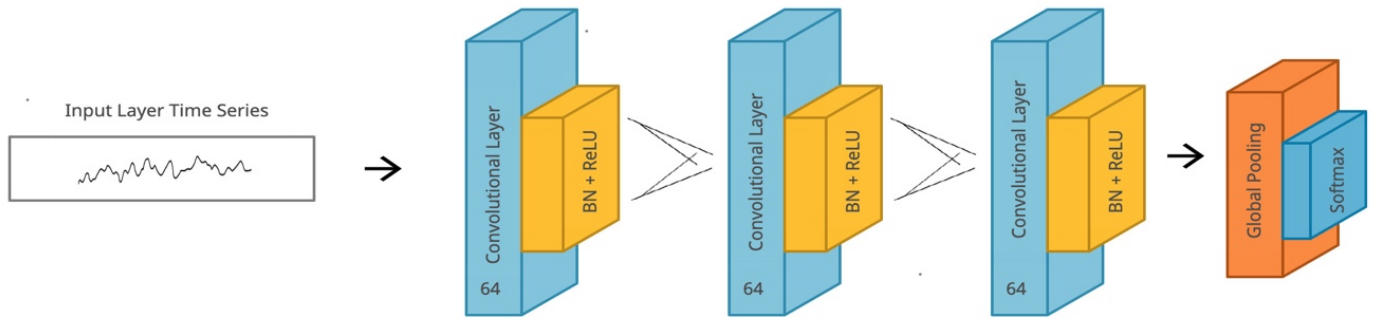

Figure 1: The Architecture of the Fully Convolutional Network

Specific applications of the FCN in health and medicine are abundant in the literature. Xu et.al. [32] apply FCNs to biomedical image segmentation and use quantization techniques to reduce overfitting. Tschandl et.al [33] incorporate the FCN in a skin lesion image segmentation model, and show performance is indistinguishable from human annotation. FCNs have recently been applied to automatic seizure detection [34,35]. Additional recent applications include lung tumor identification [36], glucose forecasting for diabetic patients [37] and cardiac diagnosis [38]. Clearly, FCNs hold promise in the health and medical domain.

## Data Augmentation for Time Series

Deep learning models tend to perform best with ample data. Data augmentation is often an effective choice for handling data scarcity and overfitting. Collecting more labeled data is not always possible, so substituting with artificially generated data becomes a viable option in such cases. Data augmentation for image classification, for example, often involves flipping, cropping, or geometric transformations of existing samples [39]. Many augmentation methods, however, do not transfer well to time series applications for the simple reason that it is not as easy to judge whether a given transformation preserves the defining features of a time series class. For instance, it is easy to determine whether a simple flipping performed on an image of a cat leaves the image identifiable as a cat. With time series, however, it is harder to evaluate at face value whether a transformation method interferes excessively with important patterns or features. Therefore, the augmentation methods reviewed here are mostly time series specific although some were adapted from areas outside of TSC.

The window slicing technique is a very widely adopted time distortion method for TSC which is largely an adaptation from computer vision [40] and was introduced by Cui et.al. [41]. This technique involves taking segments of a time series of a given class and using them as a separate series (with the same class label assigned) at training. Given a time series $T = \{t_i, t_{i+1}, \dots t_j\}$, a slice is a snipped portion of the said series:

$S_{i:j} = \{t_i, t_{i+1}, \ldots t_j\}, 1 \leq i \leq j \leq n$. If the time series is of length $n$ and the slices are of length $s$, the slicing obtains a set of $n - s + 1$ time series slices. Therefore,

$$Slicing(T, s) = \{S_{1:s}, S_{2:s+1}, \ldots, S_{n-s+1:n}\}$$

In the aforementioned method, the authors apply slicing to all time series in the training set.

Some methods fall under the category of domain-specific methods since they are suited to unique applications. Lotte [42] proposes a way to generate new data by using Fourier transforms and principal component analysis in the context of brain-machine interfaces. In a paper by Um et.al [43], multiple methods are proposed for medical sensor data including noise simulation, magnitude warping, permutations, nonlinear time warping, cropping, and multiplying the data by a random scalar.

Other augmentation methods aim to preserve the manifold of the time series, as trend and seasonality are important aspects of time series data. For example, DTW Barycenter Averaging [44] was used in a method proposed by Forestier et.al. [45] to generate a weighted average that reflects the original data's manifold.

Also relevant are resampling techniques. Resampling simply means changing the frequency of samples either by simulating a reduced sample rate on the data (downsampling) or a higher sample rate with the help of interpolation (upsampling). In a paper by Zihlmann et.al [46], electrocardiogram data is resampled for data augmentation purposes in the context of a CNN. Cao et.al [47] improve the performance of a neural network via data augmentation by concatenating ECG data, resampling, and choosing a random augmented sequence from the latter step. These augmentation methods were applied universally to the training sets. Resampling was also used by Pala et.al in training a model for sunspot forecasting [48] as an example from another domain.

# Data and Method

## Data

The dataset used in this study was the ECG200 obtained from the UCR archive [49]. Originally formatted and donated by Olszewski [50], this dataset features time series samples each representing electrical activity captured during a single heartbeat. Every sample heartbeat must belong to one of two classes: normal or myocardial infarction. The training and test set contain 100 samples each. Each time series sample has a length of 96 data points, has no missing values, and is prepended with either -1 or 1 as a class label.

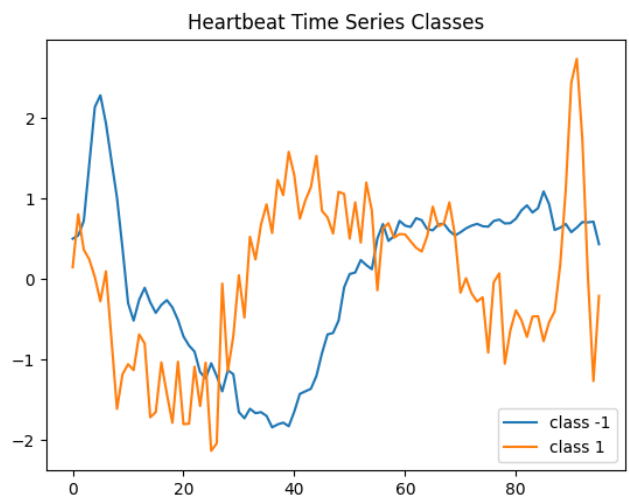

Figure 2: Heartbeat Class Time Series Examples

Note, all time series samples were already z normalized [51]. Z normalization is defined as follows: for each value in the series, the time series mean is subtracted from the said value and the difference is divided by the standard deviation. This always results in an output vector with a mean of roughly 0 and a standard deviation in a range close to 1. The other preprocessing step was to replace the class label -1 with 0, so the expected labels are 0 and 1.

## Method

The original test set was broken into two test sets referred to as *test_set_a* and *test_set_b* going forward. Both respective test sets contained an equal number of samples. In this case, *test_set_a* contained samples 1 through 50 of the original test set, and *test_set_b* contained samples 51 through 100. *Test_set_a* was incorporated in the model building process while *test_set_b* served as the true test set for the final model to be evaluated on. The training set remained unmanipulated.

The proposed method is organized into the following steps:

1. A FCN model was trained on the training set to build the original model.

2. The resulting model was used to make a separate prediction on each sample in *test_set_a*. Of these, the samples which yielded the lowest confidence predictions were selected. Specifically, an absolute difference was generated from the values in the list of probabilities for each sample (recalling that the classification function returns a probability distribution over the classes that sums to 1). For instance, if the classifier returns a probability distribution of [0.46590492, 0.5340951] over the two classes, the absolute value difference between the two values would be 0.0682. If the classifier returns a probability distribution of [0.48288137, 0.5171186] the absolute value difference would be 0.03424. Clearly, the latter example is a lower confidence prediction. The lowest confidence prediction possible is, of course, a probability distribution of [0.5, 0.5] where there is an equal chance of the sample belonging to either class. The absolute difference between the probability values in the distribution will simply be referred to as $\alpha$.

3. Low confidence samples (with a $\alpha$ value under a specified threshold) were selected. The $\alpha$ value threshold used to qualify low confidence samples for selection is itself an adjustable parameter in the new method. The results of using various $\alpha$ value thresholds at this step were separately tested and compared across various experimental models (detailed later). Note, whether the model's prediction was accurate or not was irrelevant at this step. So long as the $\alpha$ value was below the defined threshold, the sample qualified for use (the sole purpose being to extract the sample and its label for later steps).

4. Low confidence samples were compiled in a new dataset.

5. Once the low confidence samples were obtained, a data augmentation process was applied on each of the selected low confidence samples as follows. A window was sliced and extracted from each of the samples. The size of this window was always 70% of the time series, and the placement of the window was randomly determined. This window size was not arbitrary but was chosen to ensure the majority of patterns/features contained in the sample are still captured while still allowing for slight variation. Since all samples had 96 data points, the window length of the slice was always 67. The slice was then upsampled and interpolated via cubic spline interpolation [52] [53] such that the resulting series had the same length as the original series. The resulting series was then z normalized. This entire process was undergone twice for each low confidence sample, so two new resulting time series were generated from each sample:

## Pass One:

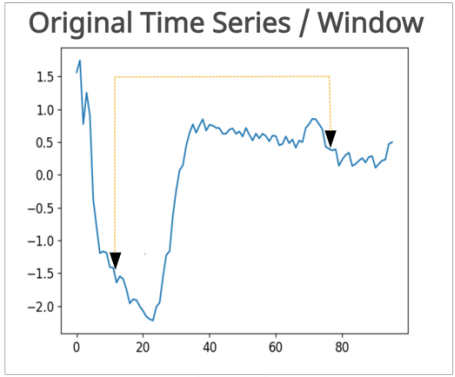

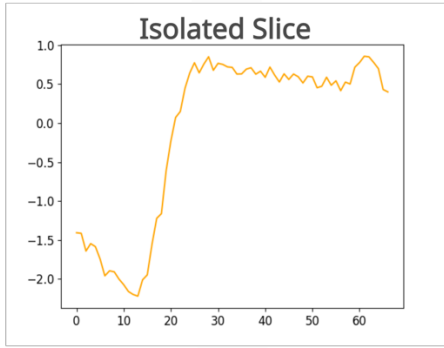

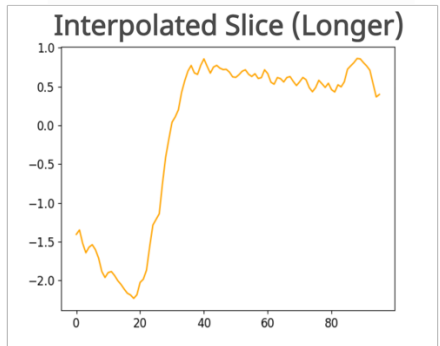

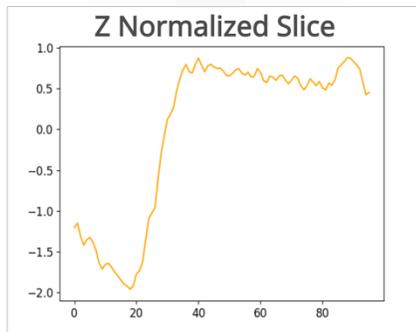

## Pass Two:

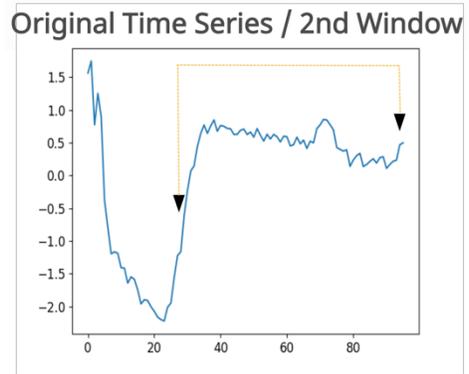

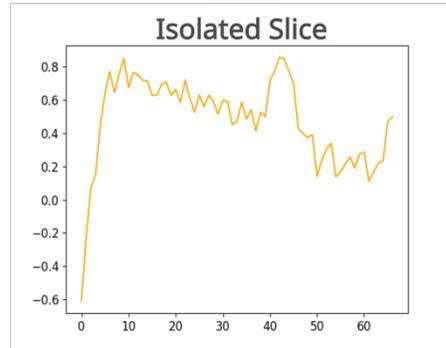

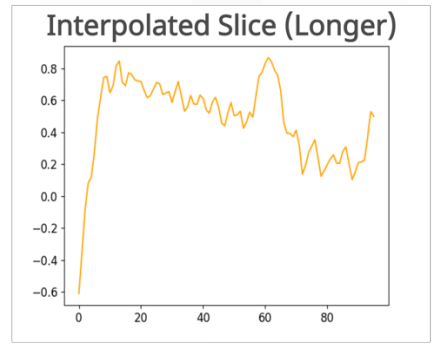

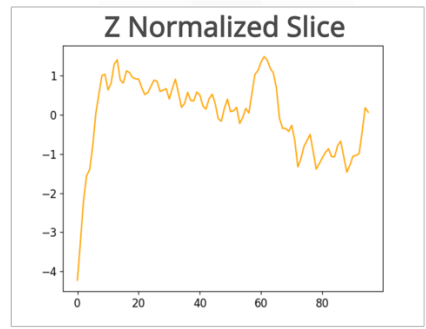

*Figure 3: Full slicing, interpolation, and normalization process on one example time series. The output of the data augmentation process on each time series is a pair of z normalized series.*

6. The newly obtained z normalized series (with their original class labels preserved) were merged with the original training set so that the newly resulting training set would contain the newly augmented samples in addition to the original training samples.

7. The FCN model was retrained from scratch on the new expanded training set.

8. The final accuracy/loss of the newly retrained model was evaluated on *test_set_b*. The newly retrained model is the final result of the process. In summary, a full map of the process is as follows:

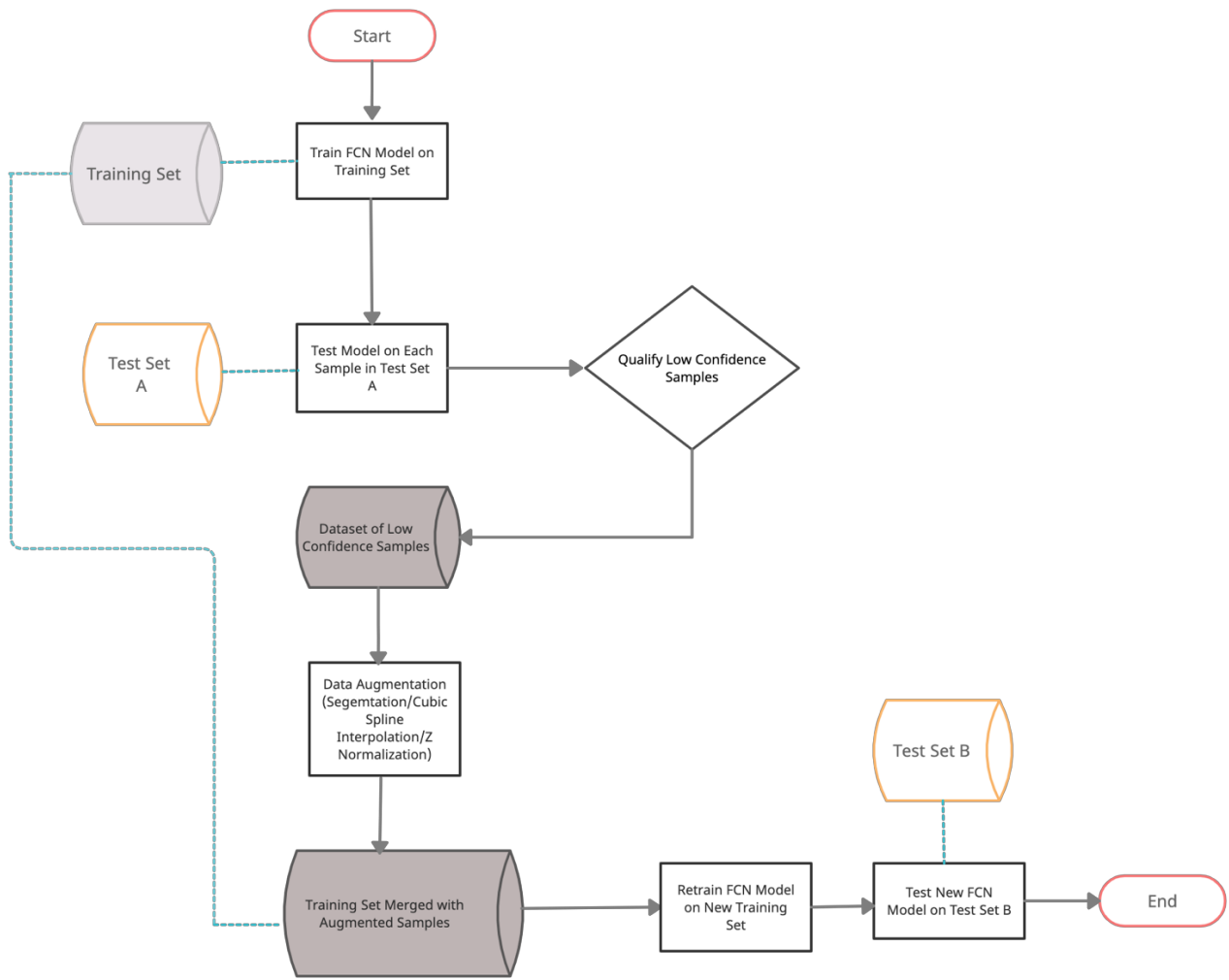

*Figure 4: A map of the new process*

The hyperparameters used for training the FCN model in the baseline as well as in the model training phases of the new method (steps 1 and 7) were strictly identical. Models were developed using Keras [54], which runs on top of TensorFlow [55]. For the sake of reproducibility, all hyperparameters used are described as follows.

Just as outlined in *Figure 1*, the FCN always had an input layer, three blocks (each consisting of the convolutional layer, batch normalization, and the ReLU activation layer), the global pooling layer, and the softmax layer. Each convolutional layer had 64 output filters. The kernel size (i.e. the length of the convolution window) for each convolutional layer was always set to 3. Dimensions of the input were preserved in the output across layers by setting padding to "same" [56]. The loss function setting used was sparse categorical crossentropy [57,58], which calculates the crossentropy loss [59] between labels and predictions to quantify model performance. There were 500 training epochs and a batch size of 32 was used.

Additional settings worth noting were best model saving and learning rate reduction. The first simply saves the best model in terms of a specified metric [60], in this case validation loss. The second was implemented to reduce the learning rate when a given metric stops improving, a functionality called ReduceLROnPlateau [61] in Keras. This was set so that the learning rate was reduced by a factor of 0.5 (with a lower bound of 0.0001 on the learning rate) if 20 epochs pass without any validation loss improvement.

For the baseline method, the FCN model was simply trained on the original training set, and accuracy/loss evaluated on *test_set_b* without the intermediate steps. Next, 8 separate experimental models were built from scratch under the new method, each with a different $\alpha$ value used as the upper limit for qualifying low confidence samples at step 3. The $\alpha$ value threshold was the only parameter that differed across tests. The resulting accuracy/loss from each model was evaluated on *test_set_b* and compared. Experiments were structured to ascertain whether models built under the new method can outperform the baseline as well as determine the optimal $\alpha$ value to use. $\alpha$ values ranged from 0.1 to 0.8 (0.8 being near the maximum observed absolute difference between probability values).

## Results

The baseline had an accuracy of 82% and a loss value of 0.4115. The results indicate using the new method influences accuracy and loss. The following table details the specific performance of the baseline and of new models built using various $\alpha$ value thresholds:

| Model | $\alpha$ Value Threshold | Accuracy | Loss |
| --- | --- | --- | --- |
| Baseline | N/A | 82% | 0.4115 |
| Model 1 | 0.1 | 88% | 0.3343 |
| Model 2 | 0.2 | 88% | 0.3337 |
| Model 3 | 0.3 | 86% | 0.354 |
| Model 4 | 0.4 | 88% | 0.3512 |
| Model 5 | 0.5 | 90% | 0.3172 |
| Model 6 | 0.6 | 84% | 0.3592 |
| Model 7 | 0.7 | 78% | 0.4288 |
| Model 8 | 0.8 | 86% | 0.3873 |

*Table 1: The Accuracy and Loss Results of the Various Models Tested*

The following charts depict the accuracy and loss of each model built under the new method:

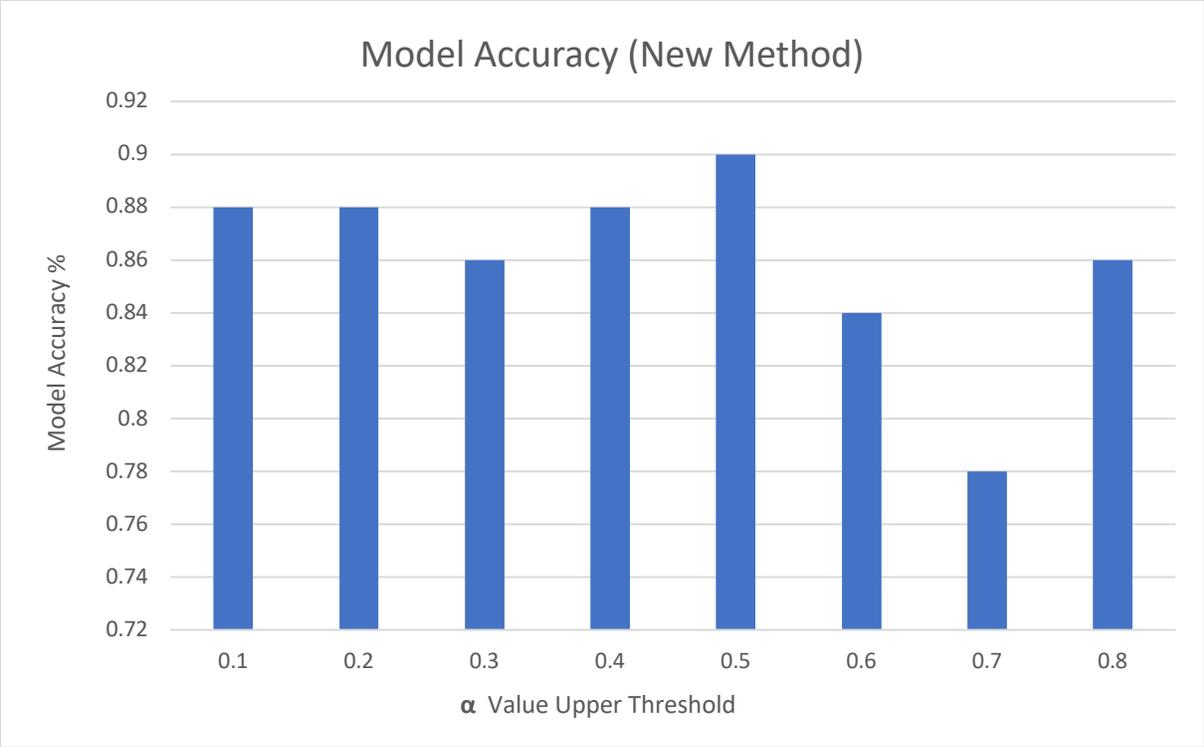

*Figure 5: Performance of All Models Built Under the New Method Displayed in Terms of Accuracy*

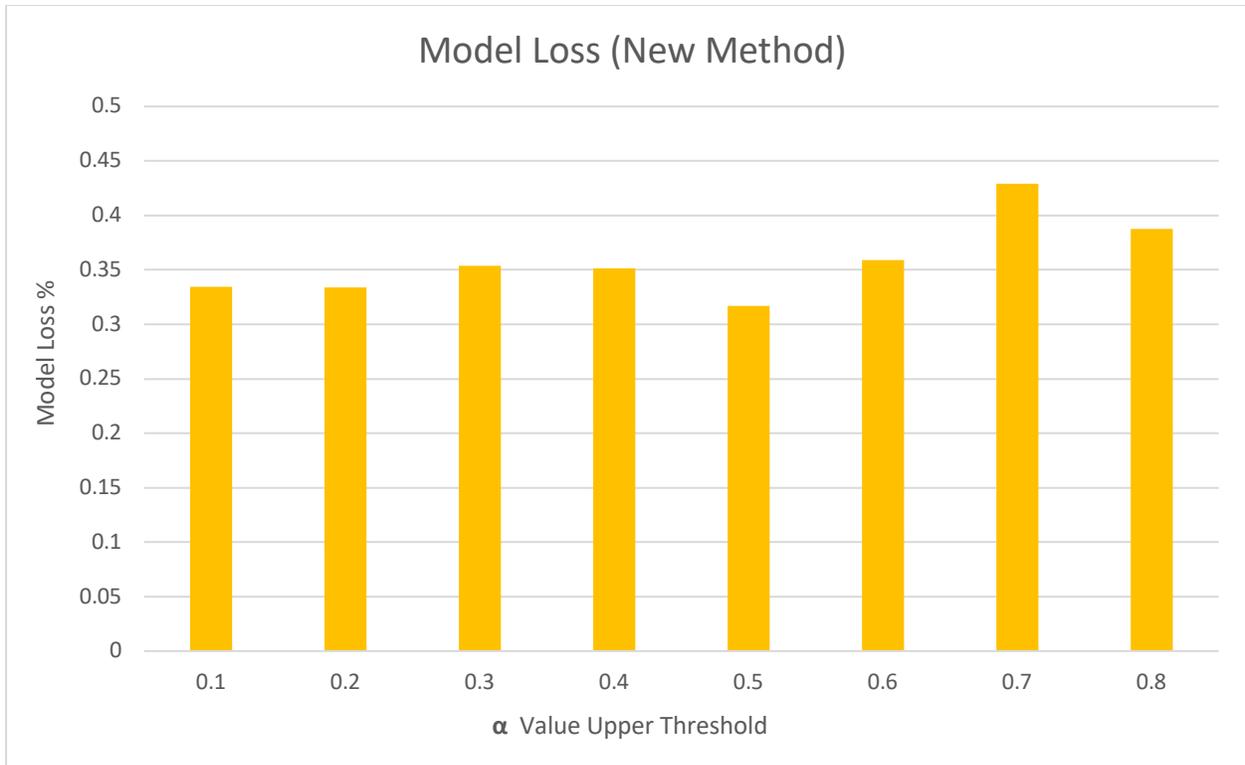

*Figure 6: Performance of All Models Built Under the New Method Displayed in Terms of Loss*

## Discussion

The experimental results suggest the new method presents tangible advantages in terms of both loss and accuracy. All models trained under the new method (except for model 7) had more optimal performance than the baseline in terms of accuracy and loss. The best performing model was model 5, achieving 90% accuracy and a loss value of 0.3172. Since the baseline had 82% accuracy and a loss value of 0.4115, this is a promising improvement.

The original speculation was that introducing augmented segments of low confidence samples into the training set could boost the model's sensitivity to less prominent features/patterns thereby improving overall performance. Increasing the upper threshold for qualifying the said low confidence samples would allow a greater number of samples to be augmented until the selected samples are not especially low confidence anymore. If the original premise was correct, it would be expected that increasing the confidence level threshold past a certain point would therefore negate the benefits of the new model, which is supposed to be selective about augmenting the low confidence samples only. The results do in fact seem to align with the premise, with the highest performance obtained by using an $\alpha$ value threshold of 0.5; performance decreases again once the $\alpha$ value threshold is increased further. In fact, out of all models built under the new method, the worst-performing ones all used $\alpha$ value thresholds greater than 0.5. This seems to support the original premise for the proposed approach.

Relating back to the work of Cao et.al [46] and Zihlmann et.al [47], machine learning clearly has strong utility in the domain of ECG data classification and can generally benefit from data augmentation techniques. These results build upon existing work by demonstrating potential validity to the technique of targeting low confidence samples for data augmentation in the context of ECG data classification. Within the limitations of this study, no claim is made regarding the benefits of the proposed approach outside the specific ECG data application.

## Conclusion

In this study, a data augmentation approach for FCNs applied to heartbeat classification was explored. The approach involved the segmentation of low confidence samples, which were reintroduced into the training set before retraining the model. The results were compared to those of a conventional implementation of the FCN. Models built under the new method showed higher performance than the baseline in terms of accuracy and loss in general (with the exception of one model). An optimal confidence level threshold for selecting low confidence samples for augmentation was also identified for this specific application. Future direction for this work includes testing the applicability of the proposed approach on other datasets in the health domain and beyond, as well as with architectures other than the FCN. In this study, a fixed window size for segmentation was used, namely 70% of the original time series length. This is a parameter that lends itself to further experimentation. Future work can also address specific methods for calculating optimal parameters inherent to the proposed approach in conjunction with traditional hyperparameter optimization.

## Abbreviations

**Time Series Classification:** TSC

**Electrocardiogram:** ECG

**Myocardial Infarction:** MI

**Dynamic Time Warping:** DTW

**Convolutional Neural Network:** CNN

**Fully Convolutional Network:** FCN

# Declarations

## Ethics Approval and Consent to Participate

The heartbeat time series data was obtained from the cited publicly available data source and has commonly been used for machine learning research. The data are anonymized with no possibility of deriving personal details. No new personal data was collected for the purpose of this study.

## Consent for Publication

The author consents to the publication of this manuscript.

## Availability of Data and Materials

All data, materials, and code used may obtained from the corresponding author upon reasonable request.

## Competing interests

The author declares no competing interests.

## Funding

This research did not receive any specific grant from funding agencies in the public, commercial, or not-for-profit sectors.

## Author's Contributions

LJ developed the approach proposed in this paper, designed the experiments, built the models, and authored the manuscript.

## Acknowledgements

The author extends thanks to Prof. Eamonn Keogh and those who undertook the colossal task of assembling the UCR time series classification archive, as well as the anonymous reviewers of this manuscript. Gratitude is also extended to Martin and Lee Budde for the sanctuary they provided during the pandemic, without which this work would not have been possible.